\documentclass{ijisa}
\usepackage[utf8]{inputenc}
\usepackage[english]{babel}
\usepackage{listings}
\usepackage{upgreek}

\usepackage{amsmath, mathrsfs}
\DeclareMathOperator*{\argmax}{arg\,max}

\title{Deep Convolutional Neural Network for Non-rigid Image Registration}

\author{Eduard F. Durech}
\affil{Simon Fraser University, Burnaby, Canada}
\affil{Email: EDurech@SFU.ca}

\thisstartpage{01}
\thisendpage{03}

\begin{document}

\abstract{Images taken at different times or positions undergo transformations such as rotation, scaling, skewing, and more. The process of aligning different images which have undergone transformations can be done via registration. Registration is desirable when analyzing time-series data for tracking, averaging, or differential diagnoses of diseases. Efficient registration methods exist for rigid (including linear or affine) transformations; however, for non-rigid (also known as non-affine) transformations, current methods are computationally expensive and time-consuming. In this report, I will explore the ability of a deep neural network (DNN) and, more specifically, a deep convolutional neural network (CNN) to efficiently perform non-rigid image registration. The experimental results show that a CNN can be used for efficient non-rigid image registration and in significantly less computational time than a conventional Diffeomorphic Demons or Pyramiding approach.}

\maketitle
\thispagestyle{thefirstpage}
\cfoot{}

\section{Introduction}
Morphing of images due to transformations introduced by time, non-idealistic properties, or multi-modal acquisitions make direct comparison of images challenging. Examples of influences leading to transformations can be movement of the acquisition device, environmental influences such as temperature, or aberrations. In medical imaging, time-series data are independently transformed and some subject images are entirely different from a template image such as in Magnetic Resonance Imaging (MRI) \cite{Talairach}. The process of aligning images, either different images or the same image under different transformations, can be achieved via registration \cite{MedImReg, MRIReg}. 

Algorithmic registration methods exist but are time-consuming and computationally expensive for more complicated registration tasks. Deep Learning has shown success in similar image processing problems such as image segmentation using a deep convolutional neural network (CNN) \cite{UNet}. CNN's can be extended to develop image transformation maps, in much the same way they develop segmentation maps, for a subject image in order to transform and align it with a template.

This paper will explore using a CNN for developing the transformation maps to register medical images with emphasis on axial brain MRI scans. The purpose is to develop an efficient method for registering images which can greatly contribute to medical image quality, processing, and diagnostics.

\subsection{Registration}
Registration is the act of aligning structures within differing images. Differences may arise due to different subjects and/or different image transformations. In medical image registration, anatomical structures are aligned leading to the ability to compare against standard templates or for analyzing the impacts of disease progression over time. Beyond the direct diagnostic and clinical significance of image registration, successful and efficient registration has the ability to streamline image processing pipelines and enhance the quality of medical images by taking multiple images in series and averaging them \cite{enFaceAv}.

Image registration relies on the assumption that transformations are independently capable of morphing an image to align with another. This assumption holds true for most medical imaging but is unable to resolve overlapping of parts in an image if information of the hidden part is unavailable, such as an anatomical part of the body being hidden behind another. This limits classic registration techniques to aligning available information in a subject image, \textit{S}, to available information in a template image, \textit{T}.

The process of deriving the correct transformation to align \textit{S} and \textit{T} is considered the process of image registration. Once the transformation is applied to \textit{S}, the resulting image is said to be registered to \textit{T}. Transformations, and the process of deriving them for registration, can be grouped into two forms, \textit{rigid} and \textit{non-rigid}.

\subsubsection{Rigid Registration}
Aligning images using linear transformations belongs to the class of rigid registration. Rigid registration treats the transformation as a rigid body which must preserve Euclidean distance between points. Examples of rigid transformations includes rotations and/or translations. Affine transformations are an extension of rigid transformations which preserve lines and parallelisms. Examples of affine transformations include scaling, shears, reflections, or any combination of them.

A plethora of methods exist for deriving rigid transformations used in rigid registration. A popular example for translation is phase correlation which uses peaks found in the cross-power spectrum of images. An efficient method for computing phase correlation is given by \cite{PCorr} \[(\Delta i,\Delta j) = \argmax_{(i,j)}\left\{\mathcal{F}^{-1}\left\{\frac{\textbf{S}\circ\textbf{T}^{*}}{|\textbf{S}\circ\textbf{T}^{*}|}\right\}\right\},\] with \(i, j\) representing the \(i^{th}\) row and \(j^{th}\) column of an image, \(\mathcal{F}\) the 2D Fourier transform, \(\textbf{S}=\mathcal{F}\{S\}\) the 2D Fourier transform of the subject image, \(\textbf{T}^{*}=\mathcal{F}\{T\}^{*}\) the conjugate of the 2D Fourier transform of the template image, and \(\circ\) the Hadamard product. This method easily extends to the \(n^{th}\) dimension by replacing \(i, j\) with \(n\) arguments and taking the \(n\)D Fourier transform in place of 2D. Phase correlation can further be extended to derive the scale and rotation transformation by initially converting \textit{S} and \textit{T} to log-polar.

\subsubsection{Non-rigid Registration}
As rigid and affine transformations are, respectively, limited to preserving Euclidean distances between points or lines and parallelisms, they falter when attempting to register images which have undergone non-rigid transformations. Non-rigid, also known as non-affine or elastic, transformations can be thought of as local rigid or affine transformation. Non-rigid transformations do not need to preserve any qualities and can independently shift each pixel in an image.

Difficulties arise in attempting to derive non-rigid transformations for registration as the inherent limitless characteristics of these transformations means parts may overlap or follow non-linear properties. Classic methods rely on pyramiding and resolving registration for multiple resolutions such as Diffeomorphic Demons \cite{Thirion, DiffDem}. These methods are robust but require a lot of computational time to complete as well as parameter tuning for specific cases. Deep learning methods for 3D CT registration have shown promising results in predicting the needed transformations for registration \cite{BIRNet, RegNet}; however, these methods were developed for 3D and feature large architectures which increase computational time.

\subsection{Warp Fields}
\begin{figure}[ht]
\centering
\includegraphics[width=\columnwidth]{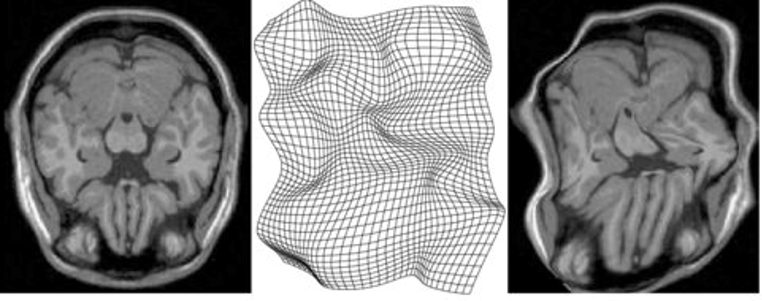} 
\caption{Non-rigid transformation example showing image, warp field, and warped image \cite{warpImage}.} \label{fig:WarpField}
\end{figure}

Non-rigid transformations can be represented by the shifts experienced by an individual or group of pixels, known as warp fields or maps. One way warp fields can be visualized is by taking a square grid and applying the transformation to it, as seen in Fig. \ref{fig:WarpField}. Mathematically, warp fields, \(\upvarphi\), can be applied to images to attain a warped image, \(W\), as \[W[i, j]=S_{i,j}\circ\upvarphi_{i,j}=S[i-\upvarphi_{i}[i, j], j-\upvarphi_{j}[i, j]].\] As some pixels will be left empty after non-rigid transformations, a common practice is to fill in empty pixels  using interpolation.

\subsection{Image Similarity}
Quantifying similarity between images is not trivial and no universal method exists. Registration techniques generally classify similarity by intensity or features. Pyramiding commonly uses normalized intensities and finds the greatest cross-correlation at each step. This method, however, is prone to "\textit{over-registering}" by aligning parts of an image whose pixels may have the same intensity but would not hold faithful to the structure of the image. An example would be if a hot-pixel were located in one item of the subject image but in another item of the template image; an ideal intensity-based registration would move the hot-pixel into the other item even if it did not belong there. Feature-based registration is possible but requires development of feature quantification algorithms, of which no universal method exists.

\subsubsection{Mean Squared Error (MSE)}
Image intensity optimizations generally minimize the mean squared error (MSE) between the warped image and template image. The error, MSE, is computed for every pixel in the image as given by \[\textrm{MSE}=\frac{1}{I\cdot J}\sum_{i,j}^{I,J}(W[i, j]-T[i,j])^2.\]

\subsubsection{Structural Similarity Index Measure (SSIM)}
One way to quantify feature similarity in images is by using their structural similarity. A popular method known as the structural similarity index measure (SSIM) computes the structural similarity between two windows, \(x\) and \(y\), as \cite{SSIM1, SSIM2} \[\textrm{SSIM}(x,y)=\frac{(2\mu_x\mu_y+c_1)(2\sigma_{xy}+c_2)}{(\mu_x^2+\mu_y^2+c_1)(\sigma_x^2+\sigma_x^2+c_2)},\] where \(\mu_{x,y}\), \(\sigma_{x,y}^2\) represent the mean and variance of \(x\) and \(y\), respectively, \(\sigma_{xy}\) represents the covariance of \(x\) and \(y\), and \(c_{1,2}\) are parameters usually defined as the weighted dynamic range of pixel values.

\begin{figure}[ht]
\centering
\includegraphics[width=.7\columnwidth]{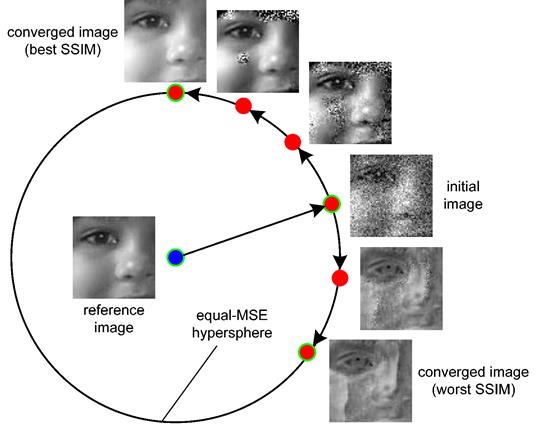} 
\caption{Reference image and images of varying SSIM quality with equal MSE \cite{SSIMcite}.} \label{fig:MSESphere}
\end{figure}

SSIM has an advantage over MSE in defining how "similar" images are. As MSE cannot distinguish between more than intensity at a given pixel, it can give the same error for source images which are both similar and very different from a template image. An example of this is displayed in Fig. \ref{fig:MSESphere} which shows a hypersphere of different images around a reference image of equal-MSE. It can be seen that images of varying similarity to the reference image can have the same MSE, but SSIM can distinguish structural similarities and properly assigns a high SSIM for similar images and a low SSIM for differing ones.

\subsection{Autoencoders}
Autoencoders are a type of deep neural network (DNN) which encode a higher-dimensional input into a lower-dimensional latent space and then decode them back to a higher-dimension. Their supporting theory is given  by the encoder filtering redundant information and representing the input in the lower-dimensional latent space which contains the most critical information about the data. The decoder will then use this critical information to generate a higher-dimensional representation of the data.

Convolutional autoencoders are common to use for deep learning-based approaches to image processing such as image segmentation. One well-known example is U-Net \cite{UNet} which uses an autoencoder structure while concatenating the outputs of the encoding levels with their respective decoding levels. The same method can be used for generating warp fields for registration instead of segmentation maps \cite{BIRNet,RegNet}.

\section{Methods}
\begin{figure}[htb]
\centering
\includegraphics[width=\columnwidth]{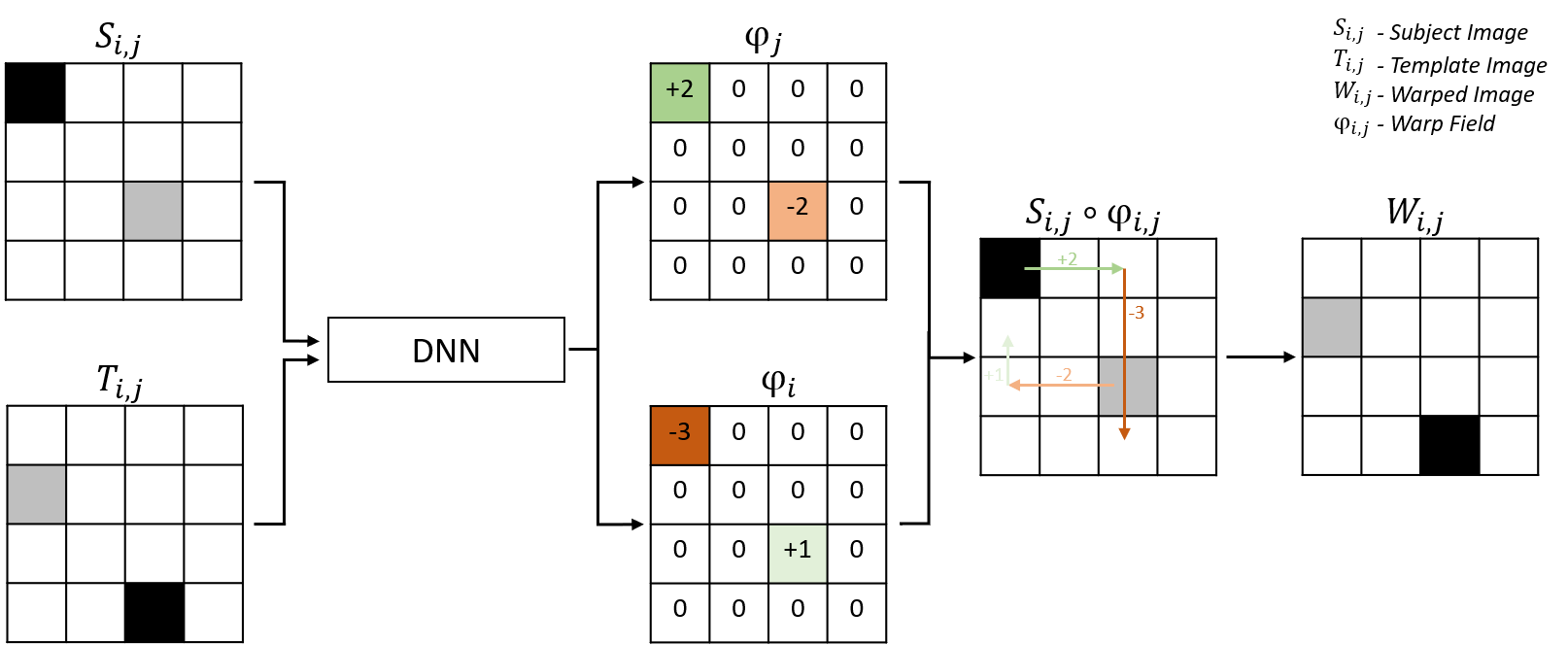} 
\caption{General pipeline.}
\label{fig:pipeline}
\end{figure}

The overall pipeline of the proposed method includes inputting the subject image, \(S\), and template image, \(T\), which are then processed and fed into a DNN. The DNN creates the warp fields, or transformation maps, \(\upvarphi_j\) and \(\upvarphi_i\), and warps \(S\) to obtain the warped image, \(W\). The warp fields were chosen to be 2D representations of the shifts for each pixel in their respective dimension, \(j\) for column or \(i\) for row. The pipeline is visualized in Fig. \ref{fig:pipeline}.

\begin{figure}[htb]
\centering
\includegraphics[width=.8\columnwidth]{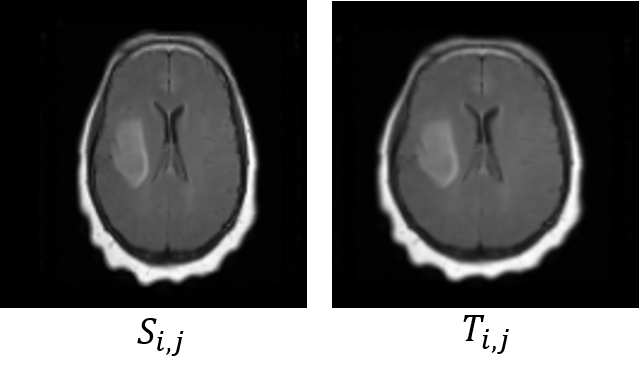} 
\caption{Example DNN input showing the normalized source image and template.}
\label{fig:input}
\end{figure}

Images were normalized before being inputted into the DNN. The input data was defined as the normalized source image and normalized template images, shown by example in Fig. \ref{fig:input}. The choice to use simply the source and template image was to not make any assumptions of how the DNN would behave. Methods such as BIRNet \cite{BIRNet} guide the DNN by inputting the difference between the subject and template as well as a Sobel-filtered subject image along with the subject image. While making assumptions about model behaviours can lead to faster training, it can also confound and restrict the ability of the model to learn. As such, minimizing the amount of information fed into a DNN can be beneficial and poses a lower risk of confusing the model.

\subsection{Deep Convolutional Neural Network (CNN)}
As convolutional autoencoders are known to be good architectures for generative image-processing tasks, a modified version of U-Net \cite{UNet} was used as the DNN. The main modifications are a change in dimensionality of the filters to distinguish higher-resolution differences in images, a 2-dimensional input of \(S_{i,j}\) and \(T_{i,j}\), a 2-dimensional output for \(\upvarphi_j\) and \(\upvarphi_i\), an additional warping step of the warp fields with \(S_{i,j}\), and a novel loss function. Hyperparameters were tuned to a learning rate of \(10^{-4}\) for Adam optimizer with a batch size of 1 trained for 100 epochs. All hidden layers used a tanh activation while the final layer used a linear activation to preserve the negative and positive values of the warp field.
\begin{figure}[htb]
\centering
\hspace*{-2mm}
\includegraphics[width=1.05\columnwidth]{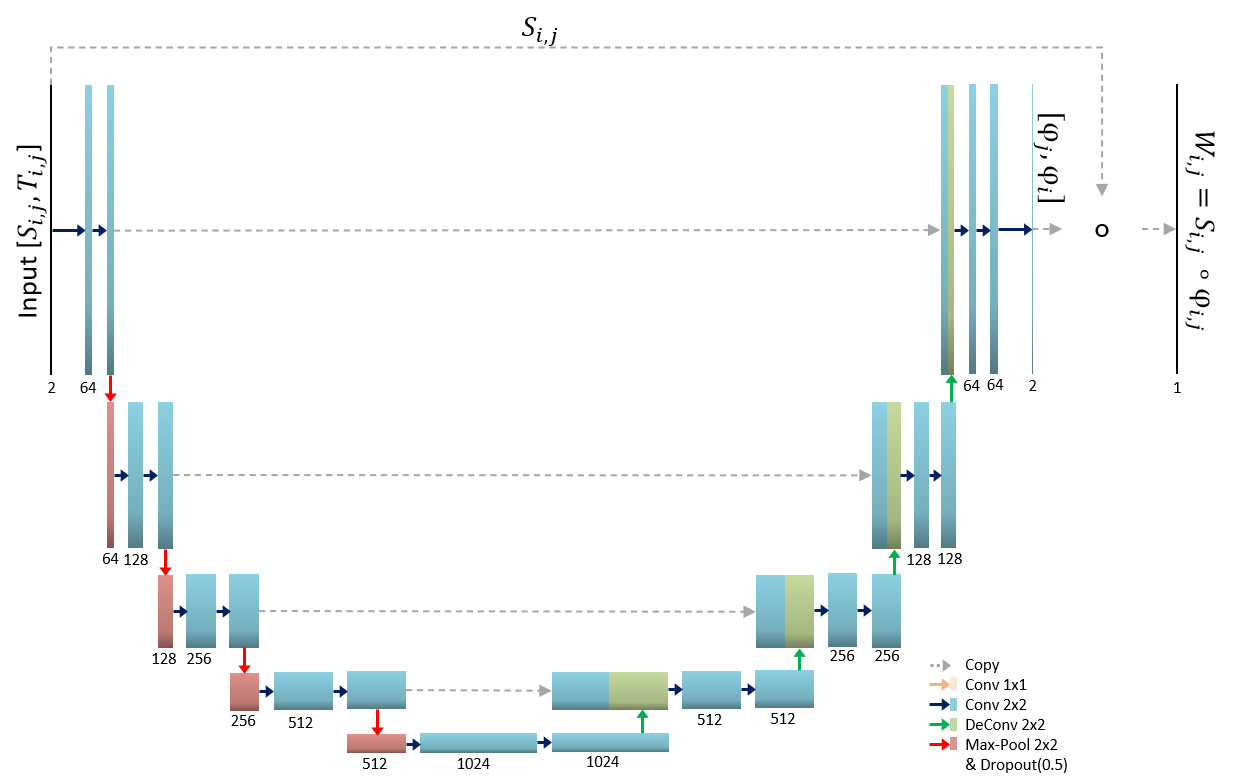} 
\caption{CNN architecture with last dimension of size \(n\) in [\(i,j,n\)] under each block.}
\label{fig:arch}
\end{figure}

\subsubsection{Weighted MSE-SSIM Loss}
The loss was computed between \(W\) and \(T\). As MSE does not capture the structural similarity between images, it was not used as the primary loss. SSIM could be used independently, but as there is no universal method of quantifying feature-based losses, a weighted addition of the two was used. As MSE should be minimized and SSIM maximized (maximal at 1), and DNN's generally seek to minimize loss, the loss was quantified as \[\textrm{Loss}=\alpha\cdot\textrm{MSE}-\beta\cdot(\textrm{SSIM}-1),\] which, in an ideal scenario with two exact images, would result in a loss of 0. The weighting coefficients \(\alpha\) and \(\beta\) were set to 10 and 1, respectively. These values were chosen to make the losses of similar magnitude for this application so both have similar influence on the loss.

Loss was not computed between warp fields as images, such as brain MRI scans, can contain many empty pixels which would be invariant to some transformations. In this case, calculating the loss using warp fields could confuse the network into learning undesirable warp fields.  This also increases the variability of solutions and thus complexity in convergence. An example of this would be rotating the outside empty region of every image with little effect on loss. As aligning the warped and template images is the ultimate goal of the network, the loss was only calculated between them.

\subsection{Data Preparation}
200 axial brain MRI slices were obtained from a TCIA public dataset \cite{TCIA} and down-sampled to 128x128 pixels. Each image underwent 5 random warps defined by linear, spherical, and sinusoidal warp fields in \(i\) and \(j\). The subsequent source image and template image (before warping) were used as the input. The warped image warped once more by its inverse field was used as the ground truth. A training/validation split of .8/.2 was used.

\section{Experiments and Results}
To test the impact of training losses on the final outcome of the model, three losses were logged for their training and validation accuracy: MSE-SSIM, SSIM (minimized as 1-SSIM), and MSE. The effects of training progression were also explored to see if the network was improving its registration over time and not just minimizing its loss.

\subsection{Losses}
\begin{figure}[t]
\centering
\hspace*{-4mm}
\includegraphics[width=1.1\columnwidth]{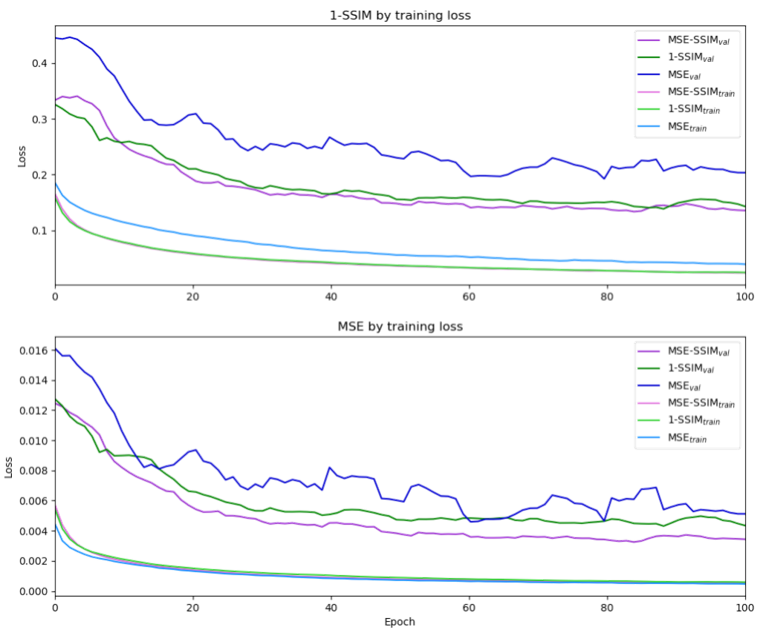} 
\caption{Training and validation 1-SSIM (top) and MSE (bottom) for MSE-SSIM, 1-SSIM, and MSE training loss over 100 epochs.}
\label{fig:lsses}
\end{figure}

A comparison of the three different training losses can be seen in Fig. \ref{fig:lsses}.
SSIM outperforms MSE in training and validation when monitoring SSIM, as is expected; however, SSIM also minimizes validation MSE greater than training with a standalone MSE loss, and is similar in training loss. This may be due to SSIM being better-suited for the problem of registration, as previously discussed. The proposed MSE-SSIM loss outperforms both a standalone SSIM or MSE loss in minimizing both validation SSIM and MSE.

\subsection{Training Progression}
\begin{figure}[!b]
\centering
\includegraphics[width=.9\columnwidth]{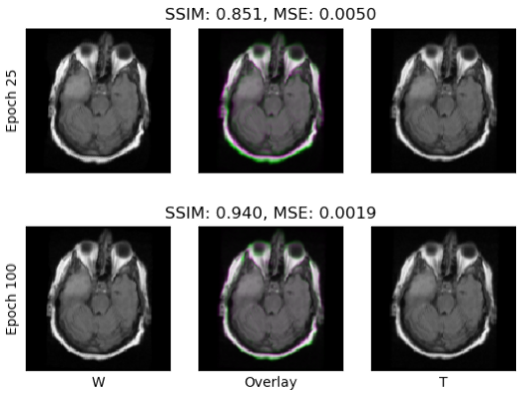} 
\caption{SSIM and MSE after 25 epochs (top) versus 100  epochs (bottom) of training. W represents the warped image, Overlay a false-colour image showing the warped image in green and template image in magenta (where an overlay results in grey), and T the template image.}
\label{fig:epch}
\end{figure}

As training progressed, the source image became more similar to the template image as shown in Fig. \ref{fig:epch}. Quantitatively, SSIM increased and MSE decreased with increasing epochs, as desired; qualitatively, the false-colour overlay shows less green and magenta and more grey, meaning the images are better-aligned after longer training. This shows the network learned some representation of non-rigid image registration.

\section{Experimental Results}
\begin{figure*}[!ht]
\centering
\includegraphics[width=.95\columnwidth]{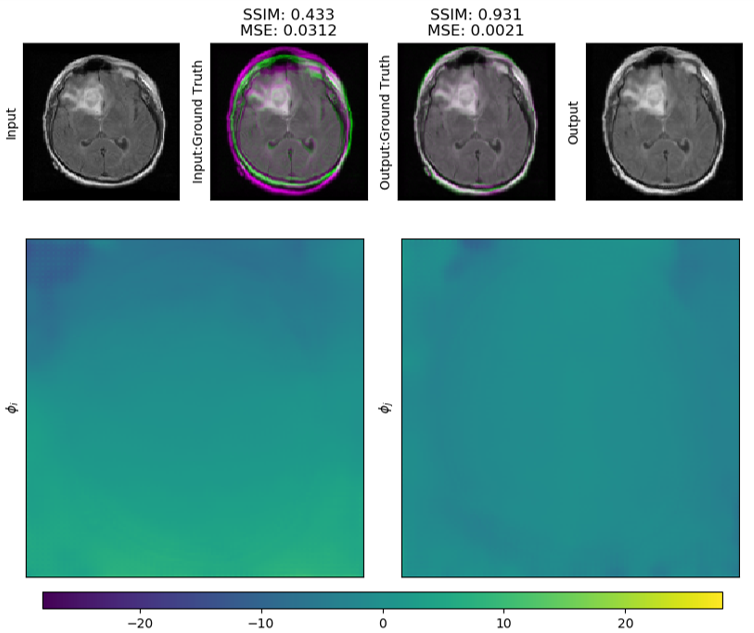} 
\includegraphics[width=.95\columnwidth]{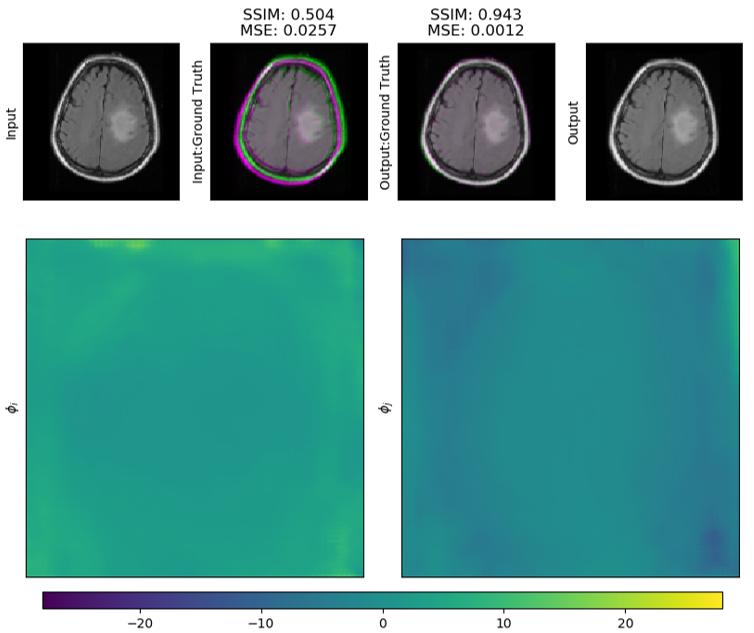} 
\includegraphics[width=.95\columnwidth]{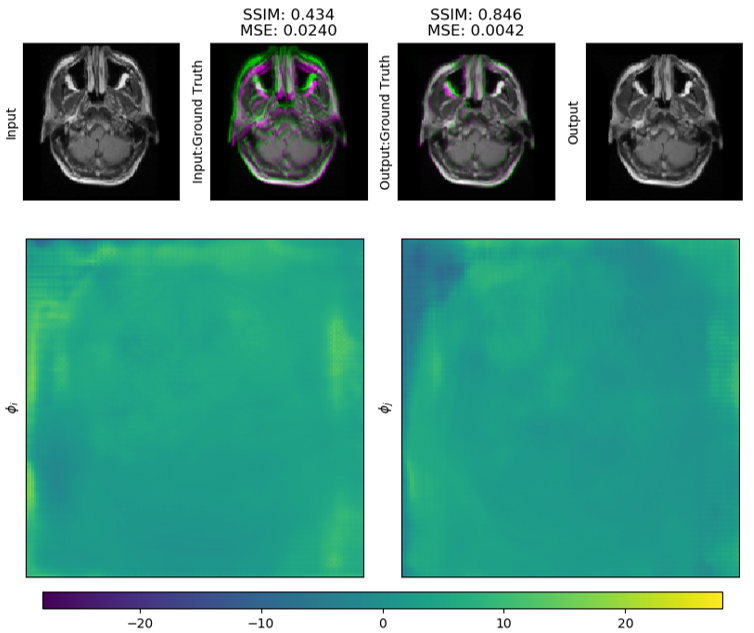} 
\includegraphics[width=.95\columnwidth]{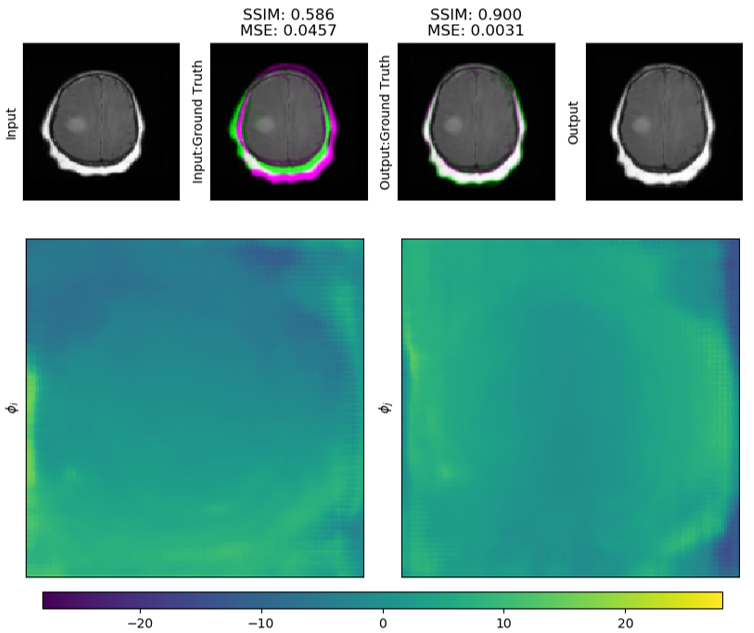} 
\includegraphics[width=.95\columnwidth]{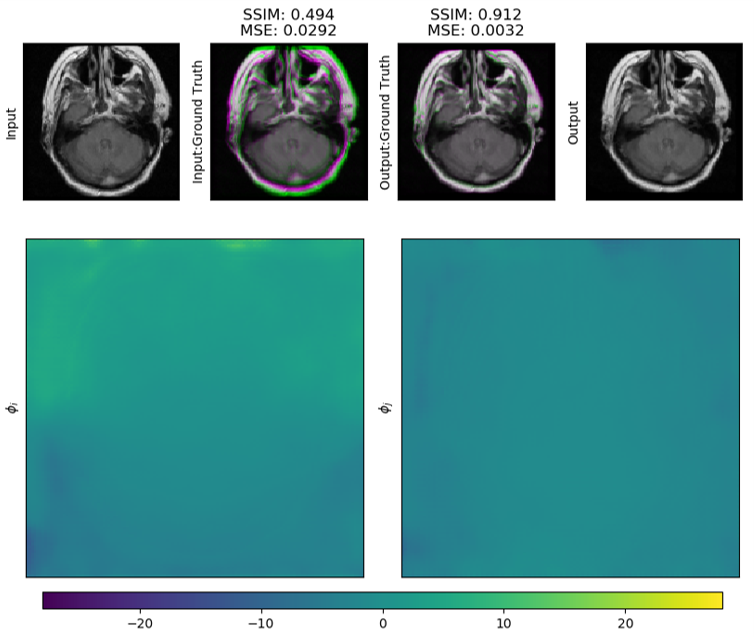}
\includegraphics[width=.95\columnwidth]{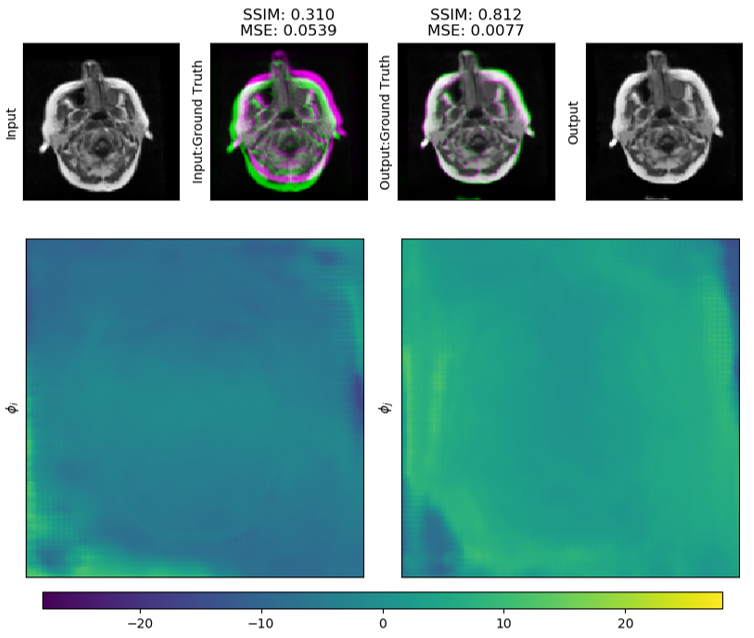}
\caption{Example results showing comparisons of inputs to outputs by SSIM, MSE, and a false-colour map of each in green versus the ground truth in magenta (grey means there is overlap). Below each example are the DNN-generated warp fields.}
\label{fig:exmpl}
\end{figure*}

This method effectively generates non-rigid warp fields to register a source image to a template as can be seen in Fig. \ref{fig:exmpl}. In the example cases, SSIM was roughly doubled while MSE was reduced by an order of magnitude. This method also efficiently generates the warp field and warps the image with an average compute time of .034s per 128x128 image taken over 100 runs.

While a lack of ground truth exists for real-world non-rigid warp fields, the synthetic data used to create these examples could be used for quantifying accuracy; however, this method of analysis suffers the same shortfalls as using the warp fields in the loss. As previously discussed, warp fields in certain regions, such as regions with empty pixels, may not impact the structural transformation of the image. Registering or calculating accuracy including these redundant regions does not give a faithful representation of the model's ability.

The warped output image, which is the desired point of reference for how well the model is performing, should be used for comparisons. A shortcoming of using the warped image is the constraint of how well we can quantify similarity between the warped and template images, as with losses. SSIM and MSE will both be used for comparisons by convention and because there is some correlation between these metrics and how well the images are aligned.

\subsection{Comparison with Algorithmic Methods}
Computational time and metric comparisons with a conventional pyramiding approach to non-rigid image registration using Diffeomorphic Demons \cite{Thirion, DiffDem} were averaged over 100 samples. The Diffeomorphic Demons algorithm was swept through an increasing amount of iterations and levels for comparison with the DNN.

\begin{figure}[!t]
\centering
\hspace*{-5mm}
\includegraphics[width=\columnwidth]{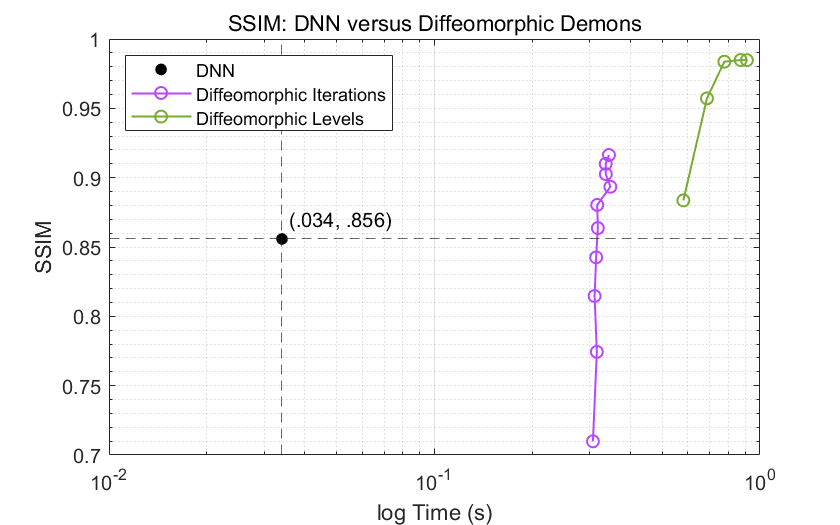}
\hspace*{-5mm}
\includegraphics[width=\columnwidth]{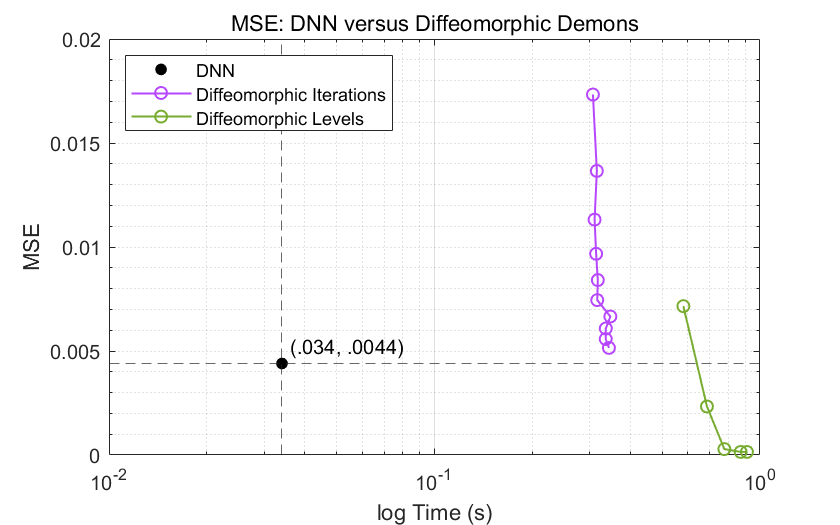} 
\caption{Comparison of DNN with Diffeomorphic Demons parameter sweeps for SSIM (top) and MSE (bottom). \newline/textit{Note: x-axes are in log-scale}.}
\label{fig:mtrc}
\end{figure}

Diffeomorphic Demons showed superior SSIM and MSE metrics under certain conditions, but took significantly longer time to compute the transformations compared to the DNN, as seen in Fig. \ref{fig:mtrc}. The DNN's metrics are as low as 15\% less than a fine-tuned Diffeomorphic Demons algorithm, but the DNN is computationally faster by at least an order of magnitude in comparison with the fastest Diffeomorphic Demons configuration.

\subsection{Probing the Decision-Making Process}
In order to understand the decision-making process of the DNN, a test image can be run through the trained network and intermediate outputs from each convolutional level can be shown.

\subsubsection{Convolutional Filters}
\begin{figure}[ht]
\centering
\hspace*{-2mm}
\includegraphics[width=1.1\columnwidth]{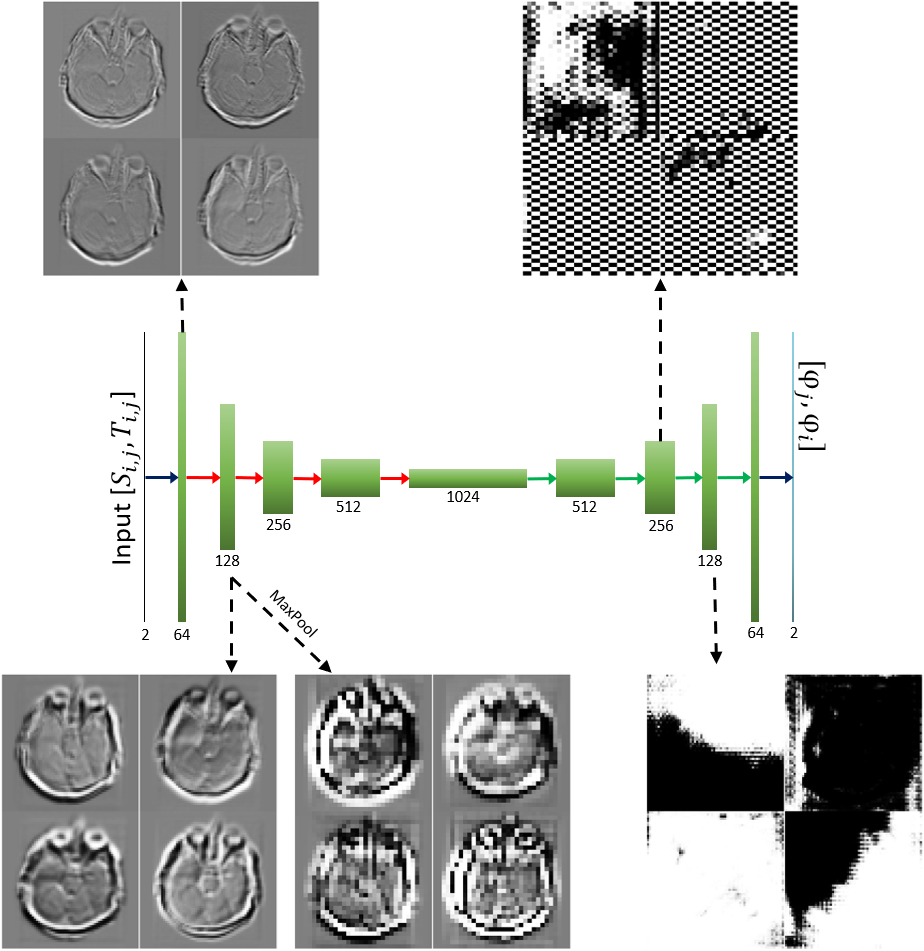}
\caption{Example intermediate outputs from convolutional levels represented as green blocks.}
\label{fig:flt}
\end{figure}

Outputs of various convolutional levels in both the encoder and decoder can be seen for a test image in Fig. \ref{fig:flt}. The initial outputs of the encoding levels resemble that of edge detectors which retain structure but dilute the amount of information into a latent space. The decoding level outputs begin to generate images which no longer represent the input, but instead that of a warp field. As the decoder uses deconvolutions, grid patterns arise while expanding the latent space.

It can be seen in Fig. \ref{fig:flt} at the 256 output-level that the network seems to be focusing on specific sub-regions to transform which become even further generalized in the 128 output-block. This is analogous to the behaviour of pyramiding by windows of varying resolutions. The network can be interpreted as a "\textit{smart}" image processor which selectively weights certain windows and regions of an image after being filtered and encoded into a constrained latent-space. This constrained latent-space features critical information such as edges and structure used in the decoding levels.

\subsubsection{Output Filter}
\begin{figure}[!hb]
\centering
\hspace*{-2mm}
\includegraphics[width=1.1\columnwidth]{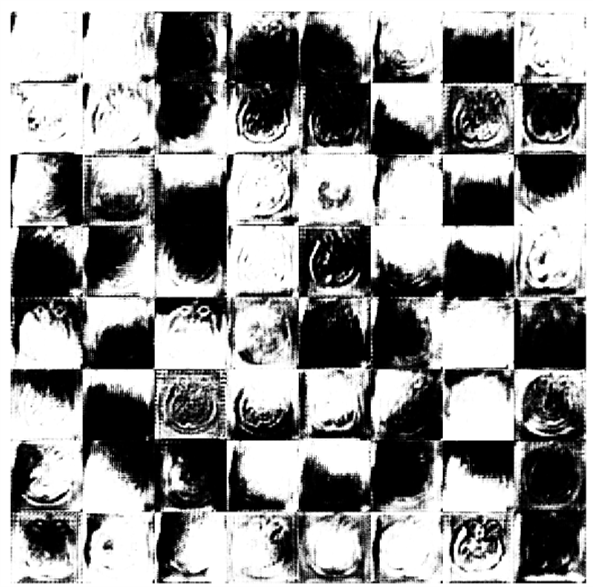}
\caption{All final filtered-outputs for a test image.}
\label{fig:otpflt}
\end{figure}

The final convolutional output layer before creating the warp field shows sub-parts used to derive the transformation. Fig. \ref{fig:otpflt} displays all outputs from this layer for a test image. Interesting patterns arise which seem to selectively isolate regions of the 2D image. These isolated regions are convolved once more to derive the final two warp maps. Some patterns show structures and edges likely used for aligning hard structure while the majority of outputs appear to be rotational isolations of the image. These likely represent sub-regions of spherical warp fields.

\section{Discussion}
Although a CNN approach is an order of magnitude faster than an algorithmic approach, the quality of the registration may not be sufficient for certain use-cases. In averaging of multiple images, source images should be near-perfectly aligned or else blurring will occur. This method has the capability of improving with different architectures, loss functions, hyperparameter tuning, and better training data. For applications which can suffice with an SSMI of <.85 and time is of concern, this method proves a good alternative to the fastest versions of algorithmic methods and boasts a higher SSMI in considerably less time. This can prove especially important for larger data as the computational demands of the algorithmic methods are non-linear for the same registration resolution.

To better represent the warp fields, \(\varphi_{i,j}\) can be reparametrized in polar coordinates as \(\varphi_{\rho,\theta}\). This could help resolve issues that may arise when attempting to register images which have transformations beyond the convolutional windows. Although it wasn't tested, such transformations could prove challenging to resolve using square convolution kernels of set dimensions. Another potential alleviation to this problem is circular convolutions.

\section{Conclusion}
Non-rigid image registration can be efficiently attained by DNN's and, more specifically, CNN's. They can achieve better quality versus time than algorithmic methods such as Diffeomorphic Demons. The proposed method uses a convolutional autoencoder to produce warp fields given a source and template image. The warp fields non-rigidly transform the source image to align with the template image, attaining registration. The network was guided using a weighted MSE-SSIM loss which achieved better SSIM and MSE metrics than either as a standalone loss. The network was also probed and its decision-making process visualized by means of displaying intermediate results throughout the network using a test image. It was discovered that the convolutional filters behave in a similar manner as algorithmic approaches do by pyramiding resolutions and windows for refined and gross transformations. This method can still be developed in order to improve its metrics and generality.

\section{ACKNOWLEDGMENT}
	I would like to thank my Professor, Dr. Ivan Bajic, for the knowledge given about deep learning systems in ENSC 413.

\bibliographystyle{IEEEtranN}
\bibliography{DCNNRIR}

\onecolumn
\lstset{numbers=left}
\section{APPENDIX A Tensorflow Python Losses}
\begin{lstlisting}
def SSIM_loss(T, S):
	return tf.reduce_mean(tf.image.ssim(T, S, 1.0))

def MSE_SSIM(T, S, alpha=10, beta=1):
	return alpha*MSE(T, S) - beta*(SSIM_loss(T, S)-1)
\end{lstlisting}

\section{APPENDIX B  Keras Python CNN Architecture}
\begin{lstlisting}
ModelIn = Input((128,128,2))

S = Lambda(lambda x: x[:,:,:,0:1])(ModelIn)

#Left Layer 1
c1 = Conv2D(64, 3, activation='tanh', padding='same')(ModelIn)
c2 = Conv2D(64, 3, activation='tanh', padding='same')(c1)
dr1 = Dropout(0.5)(c2)

#Left Layer 2
p1 = MaxPooling2D((2, 2))(dr1)
c3 = Conv2D(128, 3, activation='tanh', padding='same')(p1)
c4 = Conv2D(128, 3, activation='tanh', padding='same')(c3)
dr2 = Dropout(0.5)(c4)

#Left Layer 3
p2 = MaxPooling2D((2, 2))(dr2)
c5 = Conv2D(256, 3, activation='tanh', padding='same')(p2)
c6 = Conv2D(256, 3, activation='tanh', padding='same')(c5)
dr3 = Dropout(0.5)(c6)

#Left Layer 4
p3 = MaxPooling2D((2, 2))(dr3)
c7 = Conv2D(512, 3, activation='tanh', padding='same')(p3)
c8 = Conv2D(512, 3, activation='tanh', padding='same')(c7)
dr4 = Dropout(0.5)(c8)

#Bottom Layer
p4 = MaxPooling2D((2, 2))(dr4)
c9 = Conv2D(1024, 3, activation='tanh', padding='same')(p4)
c10 = Conv2D(1024, 3, activation='tanh', padding='same')(c9)

#Right Layer 4
uc4 = Conv2DTranspose(512, (2, 2), padding='same', strides=(2,2))(c10)
uc4 = concatenate([uc4,c8], axis=-1)
cu8 = Conv2D(512, 3, activation='tanh', padding='same')(uc4)
cu7 = Conv2D(512, 3, activation='tanh', padding='same')(cu8)

#Right Layer 3
uc3 = Conv2DTranspose(256, (2, 2), padding='same', strides=(2, 2))(cu7)
uc3 = concatenate([uc3,c6], axis=-1)
cu6 = Conv2D(256, 3, activation='tanh', padding='same')(uc3)
cu5 = Conv2D(256, 3, activation='tanh', padding='same')(cu6)

#Right Layer 2
uc2 = Conv2DTranspose(128, (2, 2), padding='same', strides=(2, 2))(cu5)
uc2 = concatenate([uc2,c4], axis=-1)
cu4 = Conv2D(128, 3, activation='tanh', padding='same')(uc2)
cu3 = Conv2D(128, 3, activation='tanh', padding='same')(cu4)

#Right Layer 1
uc1 = Conv2DTranspose(64, (2, 2), padding='same', strides=(2, 2))(cu3)
uc1 = concatenate([uc1,c2], axis=-1)
cu2 = Conv2D(64, 3, activation='tanh', padding='same')(uc1)
cu1 = Conv2D(64, 3, activation='tanh', padding='same')(cu2)

Phi = Conv2D(2, (1, 1), activation='linear', padding='same')(cu1)

W = Lambda(lambda args: tf.contrib.image.dense_image_warp(*args))([S, Phi])

FullModel = Model(inputs=ModelIn, outputs=W)
FullModelcompile(optimizer=Adam(lr=1e-4), loss='MSE_SSIM')
FullModel.fit(np.concatenate((x,y), axis=-1), y, batch_size=1, epochs=100, \
	     validation_split=.2)

\end{lstlisting}

\end{document}